\documentclass[12pt]{article}
\usepackage{amsmath,epsfig}
\usepackage{natbib}
\usepackage{color}
\usepackage{ulem}
\usepackage{verbatim}
\usepackage{graphicx}
\usepackage{rotating}
\usepackage{verbatim}
\usepackage{amssymb}

\newtheorem{model}{Model}
\newtheorem{definition}{Definition}
\newtheorem{theorem}{Theorem}
\newtheorem{corollary}{Corollary}
\newtheorem{condition}{Condition}
\begin{document}
\renewcommand{\em}{\it}
\renewcommand{\Pr}{\mathbf{P}}
\newcommand{\bm}[1]{\text{\mbox{\boldmath$#1$}}}
\baselineskip=20pt

\title{{Compound $p$-Value Statistics for Multiple Testing Procedures}}
\author{Joshua D. Habiger{\thanks{ Corresponding author: jhabige@okstate.edu.  Department of Statistics, Oklahoma State University, 301 MSCS building, 74078-1056} } and Edsel A. Pe{\~n}a
\thanks{Department of Statistics, 216 LeConte College,
University of South Carolina, 29208}}
\maketitle

\begin{abstract} Many multiple testing procedures make use of the $p$-values from the individual pairs of hypothesis tests, and are valid if the $p$-value statistics are independent and uniformly distributed under the null hypotheses.  However,
it has recently been shown that these types of multiple testing procedures are inefficient
since such $p$-values do not depend upon all of the available data.  This paper provides tools for constructing
\textit{compound} $p$-value statistics, which are those that depend upon all of the available data, but still satisfy the
conditions of independence and uniformity under the null hypotheses. As an example, a class of compound $p$-value statistics for testing for location shifts is developed.  It is demonstrated, both analytically and through simulations, that multiple
testing procedures tend to reject more false null hypotheses when applied to these compound $p$-values rather than the usual $p$-values, and at the same time still guarantee the desired type I error rate control.  The compound $p$-values, in
conjunction with two different multiple testing methods, are used to analyze a real microarray data set.  Applying either
multiple testing method to the compound $p$-values, instead of the usual $p$-values, enhances their powers.
\end{abstract}

\section{Introduction}

High throughput technology, such as the microarray, allows for thousands of pairs of hypotheses to be tested simultaneously.
The usual strategy, when testing a single pair of hypotheses, is to maximize the probability of correctly rejecting a null
hypothesis while at the same time ensuring that the probability of erroneously rejecting the null hypothesis, the type I
error rate, is controlled at some prespecified level.  However, when testing $M>1$ pairs of hypotheses simultaneously, an
additional layer of complexity arises.

Simply controlling the type I error rate at level $\alpha$ for each individual test can lead to an unpalatable number of
type I errors, especially when $M$ is large.  To combat this phenomenon, a multiple testing procedure can be used to control a globally defined error rate, such as the Family Wise Error Rate (FWER), which is the probability of committing one or more type I errors, or the False Discovery Rate (FDR), defined as the expected proportion of type I errors among rejected null hypotheses.  For a discussion of these and other global type I error rates see \citet{BenHoc95, Sto02, Sar07}.  See also \citet{WesYou93,DudVan08,Dud03} for a comprehensive review of multiple testing
methods.

Many multiple testing procedures have been developed based on the premise that data $X_m$ for testing the null hypothesis
$H_{m0}$ against the alternative hypothesis $H_{m1}$ has been ``efficiently'' reduced to some one-dimensional test
statistic, say $T_m(X_m)$, for each of the $m=1, 2, ..., M$ pairs of hypotheses. For example, methods in \citet{BenHoc95,
BenHoc00, BenKri06, GenWas04, GenWas06, Gen06, Hol79, Hom88, Hoc88, Sim86, Sid67, Sto02, StoTaySie04} make use of the
$p$-value statistics, while methods in \citet{EfrTib01, Efr08, SunCai07, JinCai07} make use of
$Z$-value statistics, which are transformed test statistics that have a standard normal distribution under the null
hypotheses.

This paper provides an answer to the question: ``How can test statistics for these multiple testing procedures be computed
in a more efficient manner, yet still allow for the procedures to be valid?''  Since many multiple testing procedures depend upon the $p$-value statistics, and are valid if they are mutually independent and uniformly distributed under the null
hypotheses, we focus on $p$-value statistics satisfying these independence and uniformity conditions. In particular, we provide tools for constructing \textit{compound} $p$-value statistics, which are those that depend upon all of the available data
$\bm{X}=(X_1, X_2, \ldots, X_M)$ via $P_1(\bm{X}), P_2(\bm{X}), \ldots, P_M(\bm{X})$, that are independent and uniformly
distributed under the null hypotheses.  As an example, we develop compound $p$-value statistics for testing for shifts in
location, and show that they satisfy the uniformity and independence conditions.  It is shown analytically and through simulations that multiple testing procedures will remain valid and tend to reject more false null hypotheses when applied to these compound $p$-values, instead of the usual \textit{simple} $p$-values, defined via $P_1(X_1), P_2(X_2), \ldots, P_M(X_M)$.

This paper proceeds as follows.  In Section 2, we present the mathematical framework and results that connect compound
$p$-value statistics to compound decision functions.  Section 3 utilizes sample-splitting ideas from \cite{Cox74} and
\cite{Rub06}, as well as results from Section 2, to develop a method for constructing compound $p$-value statistics that satisfy the independence and uniformity conditions.  Shrinkage estimators and results from Sections 2 and 3 are used to develop a class of compound $p$-value statistics for testing for location shifts in Section 4.  In Section 5, it is shown analytically and through simulation that the proposed compound $p$-value statistics, when compared to the usual simple $p$-value statistics, will lead to more powerful multiple testing procedures.  Methods are also compared to some other compound multiple testing procedures.  Compound and simple $p$-values, along with two different multiple testing procedures, are used to analyze a real microarray data set in Section 6.  The compound $p$-values allow for substantially many more rejected null hypotheses.  Some concluding remarks are in Section 7.  To make this paper more readable, all proofs of the theorems are gathered in the Appendix.

\section{Framework and Results}\label{sec2}
In this section, we present the basic framework, which was also considered in \cite{Pen11} and \cite{HabPen11a}, and establish some fundamental results that will be useful for developing compound $p$-value statistics.  Objects of main interest to us will be a random $M \times N$ matrix of observables $X = (X_{mn}, m\in\mathcal{M}, n\in\mathcal{N})\in \mathcal{X}$ with $\mathcal{M} = \{1, 2, ..., M\}$ and $\mathcal{N} = \{1, 2, ..., N\}$.  Each $X_{mn}$ need not also be 1-dimensional.  To refer to a portion of the matrix, we denote by $X[A,B] \equiv (X_{mn}:m\in A, n\in B)$. To refer to a set of columns indexed by $B \in \mathcal{N}$, we write $X[\mathcal{M},B] \equiv X[,B]$ and likewise write $X[A,]$ to refer to a set of rows.  If referring to a single column, say column $n$, we write $X[,\{n\}] \equiv X[,n]$. Similarly, we write $X[m,]$ to refer to data in row $m$.  To refer to an element of a matrix, we write $X[m,n]$.

The  distribution function of $X$ is represented by $F$.  The collection of possible distribution functions $\mathcal{F}$,
sometimes called a model for $X$, will need to be specified, such as in Model \ref{MVN}.
\begin{model}\label{MVN} Let $X\sim F\in\mathcal{F}^N$, where $ \mathcal{F}^{N} = \left\{F: F(x) = \prod_{n\in\mathcal{N}} G(x[,n];\bm{\mu},\bm{\Sigma})\right\} $ and $G(\cdot;\bm{\mu},\bm{\Sigma})$ is the multivariate normal distribution function with $M\times 1$ mean vector $\bm{\mu}$ and $M\times M$ covariance matrix $\bm{\Sigma}$.
\end{model}
This model, which assumes that columns of $X$ are independent and identically distributed according to an $M$-dimensional
multivariate normal distribution, will be considered in more detail in Section \ref{sec4}.

Pairs of hypotheses to be tested will be specified in terms of the model for the entire matrix of data.  Let
$\mathcal{F}_{m0}\subset \mathcal{F}$ and $\mathcal{F}_{m1}\subset \mathcal{F}$ be null sub-models and alternative
sub-models, respectively, such that $\mathcal{F}_{m0} \bigcup \mathcal{F}_{m1} = \mathcal{F}$ and $\mathcal{F}_{m0} \bigcap
\mathcal{F}_{m1} = \emptyset$.  The goal is to determine, for each $m\in \mathcal{M}$, which sub-model $F$ belongs to.  This is equivalent to testing the null hypothesis $H_{m0}:F\in \mathcal{F}_{m0}$ against the alternative hypothesis $H_{m1}:F\in
\mathcal{F}_{m1}$, for each $m$.

Each of the $M$ pairs of hypotheses will be tested with either a \textbf{compound} decision function, defined
$\delta_m:\mathcal{X}\rightarrow \{0,1\}$, or a \textbf{simple} decision function, defined
$\delta_m:\mathcal{X}_m\rightarrow\{0,1\}$, where $X[m,]\in \mathcal{X}_m$.
The size of $\delta_m$ is defined by
\begin{equation}\nonumber \label{size}
\eta_m = \sup_{F\in\mathcal{F}_{m0}}E_F[\delta_m(X)]
\end{equation}
where $E_F[\delta_m(X)]$ is short for $E[\delta_m(X)|X\sim F]$. Since the size $\eta_m$ of $\delta_m$ can be
specified, we write $\delta_m(\cdot;\eta_m)$.  Throughout this paper, it is assumed that for every $F\in\mathcal{F}$,
$\eta_m\mapsto\delta(x;\eta_m)$ is nondecreasing and right-continuous a.e. $[F]$.  As in \cite{Pen11} and \cite{HabPen11a}, we refer to this collection of decision functions
$\Delta_m = \{\delta_m(X;\eta_m):\eta_m\in[0,1]\}$ as a \textbf{decision process}, and refer to $\bm{\Delta} =
(\Delta_m,m\in\mathcal{M})$ as a \textbf{multiple decision process}.  Further, we say that $\Delta_m$ is compound if each $\delta_m\in\Delta_m$ is compound.

This  stochastic process framework allows for a natural definition of a $p$-value statistic.
\begin{definition} \label{P-value} The $p$-value statistic for decision process $\Delta_m = \{\delta_m(X;\eta_m):\eta_m\in[0,1]\}$ is
$P_{\Delta_m}(X) = \inf\{\eta_m\in[0,1]:\delta_m(X;\eta_m)=1\}.$
\end{definition}
Given data $X=x$, $P_{\Delta_m}(x)$ is the smallest size allowing for $H_{m0}$ to be rejected.  A $p$-value statistic is said to be \textbf{compound} if it depends on all of the data, and is written  $P_{\Delta_m}(X)$.  A $p$-value statistic will be called \textbf{simple} if it depends only on $X[m,]$, and will be written $P_{\Delta_m}(X[m,])$. Note that if a decision process is compound, then its corresponding $p$-value statistic will be compound by Definition \ref{P-value}, while if $\Delta_m$ is simple, then its $p$-value statistic will be simple.

In Theorem \ref{thmPvsD} below, we see that Definition \ref{P-value} ensures that a $p$-value statistic will be
stochastically greater than or equal to a uniform distribution under the null hypotheses.  To emphasize that this notion of uniformity depends upon the null model under consideration, we say that $P_{\Delta_m}(X)$ is $\mathcal{F}_{m0}$\textbf{-uniform} if $\sup_{F\in\mathcal{F}_{m0}}\Pr_F(P_{\Delta_m}(X)\leq t_m) = t_m$ for every $t_m\in[0,1]$, and say that the collection of $p$-value statistics $P_{\bm{\Delta}}(X) = (P_{\Delta_m}(X),m\in\mathcal{M})$ is $\mathcal{F}_{\mathcal{M}_0}$\textbf{-uniform} if $P_{\Delta_m}(X)$ is $\mathcal{F}_{m0}$-uniform for each
$m\in\mathcal{M}_0$, where $\mathcal{M}_0 = \{m:F\in\mathcal{F}_{m0}\}$ indexes those pairs of hypotheses for which $H_{m0}$ is true.
\begin{theorem}\label{thmPvsD}
The collection of $p$-value statistics $P_{\bm{\Delta}}(X) = (P_{\Delta_m}(X), m\in\mathcal{M})$ for a multiple decision
process $\bm{\Delta}$ is $\mathcal{F}_{\mathcal{M}_0}$-uniform.
\end{theorem}

Many multiple testing procedures assume that $p$-value statistics are independent of each other under the null hypotheses
and independent of $p$-value statistics from false null hypotheses.  It is therefore useful to more formally examine this
notion.  We say that $P_{\bm{\Delta}}(X)$ is $\mathcal{F}_{\mathcal{M}_0}$\textbf{-independent} if for every
$F\in\mathcal{F}_{\mathcal{M}_0}$ and $t = (t_1, t_2, ..., t_M)\in[0,1]^{M}$,
\begin{equation}\label{independentP}
\Pr_F\left(\bigcap_{m\in\mathcal{M}} [P_{\Delta_m}(X)\leq t_m]\right) =
\left[\prod_{m\in\mathcal{M}_0}\Pr_F\left(P_{\Delta_m}(X)\leq t_m\right)\right]\Pr_F\left(\bigcap_{m\in\mathcal{M}_1} [P_{\Delta_m}(X)\leq t_m]\right)
\end{equation}
where $\mathcal{M}_1 = \mathcal{M}\setminus\mathcal{M}_0$.  Likewise, the MDP $\bm{\Delta}$ is
$\mathcal{F}_{\mathcal{M}_0}$\textbf{-independent} if for every $F\in\mathcal{F}_{\mathcal{M}_0}$, $d = (d_1, d_2, ...,
d_M)\in\{0,1\}^M$, and $\eta = (\eta_1, \eta_2, ..., \eta_M)\in[0,1]^M$, we have
\begin{eqnarray}\label{independentD}
\lefteqn{\Pr_F\left(\bigcap_{m\in\mathcal{M}}[\delta_m(X;\eta_m)=d_m]\right) =}\nonumber\\
&&\left[\prod_{m\in\mathcal{M}_0}\Pr(\delta_m(X;\eta_m)=d_m)\right]\Pr_F\left(\bigcap_{m\in\mathcal{M}_1}[\delta_m(X;\eta_m)=d_m]\right).
\end{eqnarray}
Theorem \ref{thmindependent} below states that a collection of $p$-value statistics satisfy the independence condition if and only if their corresponding decision processes satisfy the condition.
\begin{theorem}\label{thmindependent}
The collection of $p$-value statistics $P_{\Delta}(X)$ for a multiple decision process $\bm{\Delta}$ is
$\mathcal{F}_{\mathcal{M}_0}$-independent if and only if $\bm{\Delta}$ is $\mathcal{F}_{\mathcal{M}_0}$-independent.
\end{theorem}
This theorem allows us to use Definition \ref{P-value} and an $\mathcal{F}_{\mathcal{M}_0}$-independent compound multiple decision process as a mechanism for defining a collection of independent compound $p$-value statistics.  The next section provides some tools for constructing this type of multiple decision process.

\section{Data Splitting}\label{sec3}
In this section, we will consider splitting one data set into two data sets via $X = (X_1, X_2)$, which we will refer to as
training data and test data, respectively.  This idea was first considered in \cite{Cox75} for testing a single pair of
hypotheses in the normal distribution setting.  \cite{Rub06} also considered sample splitting in the multiple testing
setting, but focused on a specific type of decision function for controlling the expected number of false positives.  We
avoid specifying the form of the decision function or error rate to be controlled here.  Our goal is to develop a general
$\mathcal{F}_{\mathcal{M}_0}$-uniform and $\mathcal{F}_{\mathcal{M}_0}$-independent collection of compound $p$-value statistics, which can then be used to control many different error rates.

Let $T\subset \mathcal{N}$ index a set of training data $X[,T]$ and let $\bar{T} = \mathcal{N}\setminus T$ index the set of
test data $X[,\bar T]$.  Consider decision functions taking the form $$\delta_m(X;\eta_m) = \delta_m(X[,T],X[m,\bar
T];\eta_m).$$  Note that each decision function depends on different test data $X[m,\bar{T}]$, but also depends on the same
training data $X[,T]$. Without loss of generality, we refer to the test data for $H_{m0}$ by $Z_m = X[m,\bar{T}]$ and the training
data by $Y = X[,T]$, where $Y_m = X[m,T]$.  The following independence condition will be necessary for constructing
$\mathcal{F}_{\mathcal{M}_0}$-independent $p$-value statistics.
\begin{condition}\label{condition}
The collection $\left\{(Y_m,Z_m): m\in\mathcal{M}_0\right\}$ is a mutually independent collection of random observables, and is independent of the collection  $\{(Y_m, Z_m): m\in\mathcal{M}_1\}$.
\end{condition}

We are now in a position to state Theorem \ref{mainthm}, which allows for compound $p$-value statistics to be
$\mathcal{F}_{\mathcal{M}_0}$-uniform and $\mathcal{F}_{\mathcal{M}_0}$-independent.
\begin{theorem}\label{mainthm}
Let $\bm{\Delta} = (\Delta_m,m\in\mathcal{M})$ be a multiple decision process, where $\Delta_m =
\{\delta_m(Y,Z_m;\eta_m):\eta_m\in[0,1]\}$ tests $H_{m0}:F\in\mathcal{F}_{m0}$ against $H_{m1}:F\in\mathcal{F}_{m1}$ for
each $m$.  If, for every $F\in\mathcal{F}_{m0}$, $E_F(\delta_m(Y, Z_m;\eta_m)|Y) = \eta_m$ for every $m\in\mathcal{M}_0$ and $\eta_m\in[0,1]$, then $P_\bm{\Delta}(Y,Z)$ is $\mathcal{F}_{\mathcal{M}_0}$-uniform. If, in addition, Condition
\ref{condition} is satisfied, then $P_\bm{\Delta}(Y,Z)$ is $\mathcal{F}_{\mathcal{M}_0}$-independent.
\end{theorem}
It is important to emphasize that the decision processes, and hence $p$-value statistics, are allowed to be dependent under the alternative hypotheses.  In fact, we will see that improvements over the usual simple $p$-values will be made by constructing $p$-values that are dependent under the alternative hypotheses.

\section{Composite Hypotheses}\label{sec4}

In this section we will develop compound $p$-value statistics for testing multiple pairs of hypotheses regarding location
parameters.  The strategy is to develop an $\mathcal{F}_{\mathcal{M}_0}$-independent compound multiple decision process, and then make use of Definition \ref{P-value} and Theorem 3 to derive $\mathcal{F}_{\mathcal{M}_0}$-uniform and
$\mathcal{F}_{\mathcal{M}_0}$-independent compound $p$-values.  In what follows, we utilize Model 1 to develop the $p$-values, but results are not limited to this setting.  This notion is discussed in more detail in Section 5.

Assume that $X$ has distribution function $F\in \mathcal{F}^{N}$ where $\mathcal{F}^{N}$ is Model
\ref{MVN} with mean vector $\frac{1}{N}\bm{\mu}$ and covariance matrix $\frac{1}{N}\bm{I}$.  Here, we let the mean vector
and covariance matrix depend on $N$ so that, as we will see, the distribution of the sufficient statistics for the
hypotheses of interest is free of $N$.  The pairs of hypotheses are $H_{m0}:F\in\mathcal{F}_{m0}^{N} = \{F\in
\mathcal{F}^{N}:\mu_m=0\}$ and $H_{m1}:F\in\mathcal{F}_{m1}^{N} = \{F\in \mathcal{F}^{N}:\mu_m\neq0\}$ for each $m$.  The
collection of true null hypotheses is indexed by $\mathcal{M}_0 = \{m:\mu_m=0\}$ and the collection of false null hypotheses
is indexed by $\mathcal{M}_1 = \{m:\mu_m\neq 0\}$. We simplify our notation by writing vectors of sufficient statistics for
$\bm{\mu}$ with respect to the training data $X[,T]$ and test data $X[,\bar T]$ by
$$Y = \sum_{n\in T}X[,n] \mbox { and }Z = \sum_{n\in \bar T}X[,n],$$
respectively.  Denote the vector of sufficient statistics for the complete data by
$$W = \sum_{n\in\mathcal{N}}X[,n].$$
Note that $Y\sim MVN(\lambda^2\bm{\mu},\lambda^2 \bm{I})$ and $Z\sim MVN((1-\lambda^2)\bm{\mu},(1-\lambda^2)\bm{I})$ where
$\lambda^2 = |T|/|N|$ is the proportion of training data and $1-\lambda^2$ is the proportion of test data, and $W\sim
MVN(\bm{\mu}, \bm{I})$.

To motivate our compound decision function, we first consider a simple decision function, which is allowed to depend on the
unknown $\bm{\mu}$, rather than training data $Y$, and test data $Z_m$.  It is defined via
\begin{equation}\label{ordelta}
\delta_m(\bm{\mu},Z_m;\eta_m) = I\left(\frac{Z_m}{\sqrt{1-\lambda^2}}\leq l_m(\bm{\mu},\eta_m)\right) + I\left(\frac{Z_m}{\sqrt{1-\lambda^2}}\geq u_m(\bm{\mu},\eta_m)\right)
\end{equation}
where $l_m(\bm{\mu},\eta_m) = \Phi^{-1}(\eta_m h_m(\bm{\mu}))$ and $u_m(\bm{\mu},\eta_m) = \Phi^{-1}(1-\eta_m[1-h_m(\bm{\mu})])$ are lower- and upper-tail cutoffs, respectively, $\Phi(\cdot)$ is the standard normal distribution function, and $h_m:\Re^{M}\rightarrow [0,1]$ acts as a weight governing $l_m(\bm{\mu},\eta_m)$ and $u_m(\bm{\mu},\eta_m)$.  Notice that when $\mu_m = 0$, $Z_m/\sqrt{1-\lambda^2}$ has a standard normal distribution, and hence $E_F(\delta_m(\bm{\mu},Z_m;\eta_m)) = \eta_m$ for any $h_m(\bm{\mu})$.  Since $(Z_m,
m\in\mathcal{M})$ is an independent collection, $\bm{\Delta}$ is an $\mathcal{F}_{\mathcal{M}_0}$-independent multiple
decision process.

Now, an Oracle, who knows $\bm{\mu}$, could choose $h_m(\bm{\mu})$ to maximize the power of $\delta_m$, defined via
\begin{eqnarray}\label{orpow} \nonumber
\lefteqn{\beta_m(\bm{\mu},\lambda,\eta_m) = E_F[\delta_m(\bm{\mu},Z_m;\eta_m)]}\\
&&=\Phi\left(l_m(\bm{\mu},\eta_m)-\sqrt{1-\lambda^2}\mu_m\right) + 1 - \Phi\left(u_m(\bm{\mu},\eta_m) - \sqrt{1-\lambda^2}\mu_m\right),
\end{eqnarray}
thereby maximizing the average power
\begin{equation}\label{oravepow}
\beta(\bm{\mu},\lambda,\bm{\eta}) = \frac{1}{M_1}\sum_{m\in\mathcal{M}_1}\beta_m(\bm{\mu},\lambda,\eta_m),
\end{equation}
were $M_1 = |\mathcal{M}_1|$ is the number of false null hypotheses.  It can be verified that for each $m\in\mathcal{M}_1$ and for a fixed $\lambda$ and $\bm{\eta} = (\eta_m, m\in\mathcal{M})$, $\beta_m(\bm{\mu},\lambda,\eta_m)$, and hence $\beta(\bm{\mu},\lambda,\bm{\eta})$, is maximized by
defining $h_m(\bm{\mu}) = I(\mu_m\leq 0)$. Thus, the Oracle decision function is
\begin{eqnarray*}
 \delta_m^{(or)}(\mu_m,Z_m;\eta_m) &=& I\left(\frac{Z_m}{\sqrt{1-\lambda^2}}\leq l_m^{(or)}(\mu_m,\eta_m)\right)
+ I\left(\frac{Z_m}{\sqrt{1-\lambda^2}}\geq u_m^{(or)}(\mu_m,\eta_m)\right),\end{eqnarray*}
where $l_m^{(or)}(\mu_m,\eta_m) = \Phi^{-1}(\eta_m I(\mu_m\leq0))$ and $u_m^{(or)}(\mu_m,\eta_m) = \Phi^{-1}(1-\eta_m[1-I(\mu_m\leq0)])$ are the lower-tail
and upper-tail Oracle cutoffs arising by plugging in $I(\mu_m\leq0)$ for $h_m(\bm{\mu})$ in $l_m(\bm{\mu},\eta_m)$ and
$u_m(\bm{\mu},\eta_m)$ in expression (\ref{ordelta}).  It should be noted that other optimality criterion have been considered.  \cite{Sto07} and \cite{Spj72} considered maximizing
the expected number of true positives (ETP), which can be written ETP = $M_1\beta(\bm{\mu},\lambda,\bm{\eta})$, while
\cite{Pen11} considered minimizing the expected number of ``missed discoveries'' or missed discovery rate (MDR), which can
be defined by MDR = $M_1[1-\beta(\bm{\mu},\lambda,\bm{\eta})] = M_1 - ETP$.  Both of these optimality criterion are
satisfied by maximizing $\beta(\bm{\mu},\lambda, \bm{\eta})$.

The Oracle $p$-values can be derived using Definition \ref{P-value}. Writing
$$I\left(\frac{Z_m}{\sqrt{1-\lambda^2}}\leq l_m^{(or)}(\mu_m,\eta_m)\right) = I\left(\frac{\Phi\left(\frac{Z_m}{\sqrt{1-\lambda^2}}\right)}{I(\mu_m\leq0)}\leq \eta_m\right)$$
and
$$ I\left(\frac{Z_m}{\sqrt{1-\lambda^2}}\geq l_m^{(or)}(\mu_m,\eta_m)\right) = I\left(\frac{1-\Phi\left(\frac{Z_m}{\sqrt{1-\lambda^2}}\right)}{1-I(\mu_m\leq 0)}\leq \eta_m\right),$$
with $a/0 = \infty$ for $a>0$, it follows from Definition 1 that the Oracle $p$-value statistic for $\Delta_m^{(or)} = \{\delta_m^{(or)}(\mu_m, Z_m;\eta_m):\eta_m\in[0,1]\}$  can
be written as
\begin{equation}\label{OracleP-val}
P_{\Delta_m^{(or)}}(\mu_m,z_m) =\min\left\{\frac{\Phi\left(\frac{Z_m}{\sqrt{1-\lambda^2}}\right)}{I(\mu_m\leq 0)},
\frac{1-\Phi\left(\frac{Z_m}{\sqrt{1-\lambda^2}}\right)}{1-I(\mu_m \leq 0)}\right\}.
\end{equation}
We make use of this particular expression to allow for a more straightforward comparison of the Oracle $p$-value and the
compound $p$-value, which is presented next.  It is important to note that since $\bm{\Delta}^{(or)} = (\Delta_m^{(or)}, m\in\mathcal{M})$ is an $\mathcal{F}_{\mathcal{M}_0}^N$-independent MDP, $P_{\bm{\Delta}^{(or)}}(\bm{\mu},Z) = (P_{\Delta_m^{(or)}}(\mu_m,Z_m), m\in\mathcal{M})$ is $\mathcal{F}_{\mathcal{M}_0}^{N}$-uniform and $\mathcal{F}_{\mathcal{M}_0}^{N}$-independent.

Using training data $Y$ to estimate $I(\mu_m\leq0)$ results in a \textbf{compound} decision function
\begin{equation}\label{delta1}\nonumber
\delta_m^{(c)}(Y,Z_m;\eta_m) = I\left(\frac{Z_m}{\sqrt{1-\lambda^2}}\leq l_m(Y,\eta_m)\right) + I\left(\frac{Z_m}{\sqrt{1-\lambda^2}} \geq u_m(Y,\eta_m)\right),
\end{equation}
where $l_m(Y,\eta_m) = \Phi^{-1}(\eta_m h_m(Y))$ and $u_m(Y,\eta_m) = \Phi^{-1}(1-\eta_m[1-h_m(Y)])$ are lower- and
upper-tail cutoffs, respectively, and $h_m(Y)$ estimates $I(\mu_m\leq 0)$.  Arguments similar to those made above can be used to show that the compound $p$-value statistic for $\Delta_m^{(c)}$ is
\begin{equation}\label{CompP-val}
P_{\Delta_m^{(c)}}(Y,Z_m) = \min\left\{\frac{\Phi\left(\frac{Z_m}{\sqrt{1-\lambda^2}}\right)}{h_m(Y)},
\frac{1-\Phi\left(\frac{Z_m}{\sqrt{1-\lambda^2}}\right)}{1-h_m(Y)}\right\}.
\end{equation}
See \cite{HabPen11a} for other forms of simple $p$-values for composite hypothesis testing.

Notice that given $Y=y$, if $h_m(y) = I(\mu_m\leq0)$, then the compound and Oracle $p$-value statistics are equivalent.  Hence, the goal will be to develop an $h_m(Y)$ that estimates $I(\mu_m\leq0)$ ``well''.  However, before proceeding, it is important to point out that these compound $p$-value statistics are $\mathcal{F}_{\mathcal{M}_0}^N$-independent and
$\mathcal{F}_{\mathcal{M}_0}^N$-uniform, regardless of the performance of $h_m(Y)$, and hence lead to valid multiple testing procedures.  This result is formally stated in Corollary \ref{cor}.
\begin{corollary}\label{cor}
Let $\mathcal{M}_0 = \{m\in\mathcal{M}:\mu_m = 0\} $.  Then
$P_{\bm{\Delta}^{(c)}}(Y,Z) = (P_{\Delta_m^{(c)}}(Y,Z_m), m\in\mathcal{M})$
is $\mathcal{F}_{\mathcal{M}_0}^N$-uniform and $\mathcal{F}_{\mathcal{M}_0}^N$-independent.
\end{corollary}

Next, we develop a class of estimators of $I(\mu_m\leq 0)$ using empirical Bayes ideas.  Assume, for the moment, that $\mu_m$ is random, and for $m\in\mathcal{M}$, let $J_m = I(\mu_m\neq 0)$ be independent and identically distributed Bernouli random variables with success probability $p$.  Note that if $J_m = 1$, then $H_{m0}$ is false.  Further, assume that the distribution function for $\mu_m$, given $J_m = 1$, is
\begin{equation}\label{prior}\nonumber
G(\mu_m|J_m=1;\theta,\tau) = \Phi\left(\frac{\mu_m - \theta}{\tau}\right)
\end{equation}
Since $Y_m|(\mu_m,J_m=1)\sim N(\lambda^2\mu_m,\lambda^2)$ and $\mu_m|(J_m=1)\sim N(\theta,\tau^2)$, we have that $\mu_m|(Y_m=y_m,J_m=1)$ has a normal distribution with mean $(y_m\tau^2 + \theta)/(\lambda^2\tau^2 + 1)$ and variance $\tau^2/(\lambda^2\tau^2 + 1).$  See, for example, \cite{CasBer02}, page 326.
Thus, the posterior distribution function of $\mu_m$,
given ($Y_m=y_m, J_m=1$), is
$$G(\mu_m|Y_m=y_m,J_m=1;\theta,\tau) = \Phi\left(\left[\mu_m-\frac{y_m\tau^2 + \theta}{\lambda^2\tau^2+1}\right]\sqrt{\frac{\lambda^2 \tau^2+1}{\tau^2}}\right).$$
Here we condition on $J_m=1$ since, when $J_m = 0$, $E_F[\delta_m^{(c)}(Y,Z_m;\eta_m)] = \eta_m$ regardless of $h_m(Y)$, and since the goal is to maximize the power of a $\delta_m$ when $\mu_m\neq 0$. We should not be concerned with maximizing the power of $\delta_m$ when $J_m=0$ since this would correspond to maximizing the probability of committing a type I error, i.e., making a false discovery.

Since $\theta$ and $\tau$ are not known, the estimate of $I(\mu_m\leq0)$ given by $h(y_m,\theta,\tau) =
G(0|Y_m=y_m, J_m=1;\theta,\tau)$ is not yet computable.  In an effort to develop easy-to-compute $p$-value statistics, we develop
method-of-moments (MOM) estimators for these parameters.  Still viewing ($J_m,\mu_m$) as random, we get
%
$$E(Y_m) = E(E(Y_m|J_m)) = p\lambda^2\theta$$
%
and
%
$$Var(Y_m) = E(Var(Y_m|J_m)) + Var(E(Y_m|J_m)) = \lambda^2 + \lambda^4p(\theta^2[1-p] + \tau^2).$$
%
Setting these expressions equal to the sample mean $\bar{y}$ and sample variance $s^2$ of $y_1, y_2, ..., y_M$,
respectively, and solving for $\theta$ and $\tau$ yields the MOM estimates
\begin{equation}\nonumber
\hat{\theta}(y) = \frac{\bar{y}}{\lambda^2p}
\end{equation}
and
\begin{equation}\nonumber
\hat{\tau}^2(y) = \max\left\{\frac{s^2-\lambda^2 - \bar{y}^2(1-p)/p}{p\lambda^4}, 0\right\}.
\end{equation}
Note that we set $\hat \tau^2$ equal to 0 whenever the solution yields a negative estimate of $\tau^2$.

Both of these MOM estimators now depend on the proportion of false null hypotheses $p$, and hence it is necessary to either
specify or estimate $p$.  In the next section, we will consider setting $p = 1$, and we will refer to resulting estimators
of $\theta$, $\tau$, and $I(\mu_m\leq 0)$ as approximate minimax estimators since this specification corresponds to the
assumption that all null hypotheses are false.  Other possible specification of $p$ will be considered in Section 6.  For
now, we develop a class of MOM estimators for $p$ using the fact that
\begin{equation}\label{ineq1}
E[I(-\epsilon \leq Y_m \leq \epsilon)] = (1-p)A(\epsilon) + p
B(\epsilon;\theta,\tau)
\geq (1-p)A(\epsilon),
\end{equation}
where
$$A(\epsilon) = \Pr(-\epsilon\leq Y_m\leq \epsilon|J_m=0) = \Phi(\epsilon/\lambda) - \Phi(-\epsilon/\lambda)$$ and
$B(\epsilon;\theta,\tau) = \Pr(-\epsilon\leq Y_m\leq \epsilon|J_m=1)$. Making use of expression (\ref{ineq1}) and sample moment $\frac{1}{M}\sum_{m\in\mathcal{M}}I(-\epsilon\leq y_m\leq \epsilon)$, we get
$$ \hat{p}(y;\epsilon) = 1 - \frac{1}{M}\frac{\sum_{m\in\mathcal{M}}I(-\epsilon\leq y_m \leq\epsilon)}{\Phi(\epsilon/\lambda) - \Phi(-\epsilon/\lambda)},$$
which no longer depends upon $\tau$ or $\theta$, but does depend on the tuning parameter $\epsilon$.  This type of estimator has been studied in the multiple testing literature, though not in this sample splitting setting.  See \cite{EfrTib01} or
\cite{Efr04}, for example.  For other types of estimators of $p$, see \cite{JinCai07}, \cite{Lan05}, \cite{Net06}, \cite{Sto02}, \cite{StoTaySie04}, among others. The choice of $\epsilon$ will be considered in more detail in Sections 5 and 6.

Finally, plugging
\begin{equation}\label{estimators}
\hat{\theta}(y) = \frac{\bar{y}}{\lambda^2\hat p(y;\epsilon)} \mbox{ and } \hat{\tau}^2(y) = \max\left\{\frac{s^2-\lambda^2 - \bar{y}^2[1-\hat p(y;\epsilon)]/\hat p(y;\epsilon)}{\hat p(y;\epsilon)\lambda^4}, 0\right\}
\end{equation}
for $\theta$ and $\tau$ in $G(0|Y_m=y_m,J_m=1;\theta, \tau)$ yields the estimate of $I(\mu_m\leq 0)$ given by
\begin{equation}\label{hhat}
h_{m}(y) = \Phi\left(-\frac{y_m\hat\tau^2(y) + \hat\theta(y)}{\sqrt{\hat\tau^2(y)(\lambda^2\hat\tau^2(y)+1)}}\right).
\end{equation}
In the next section, we study how the choice of $\lambda^2$ and the performance of $h_m(Y)$ affects the power of the compound and Oracle decision functions, and hence affects the performance of their corresponding $p$-value statistics.

\section{Assessment}
\subsection{Analytical Assessment}
To better understand the performance of the compound $p$-value statistic and ultimately determine how $\lambda^2$ and $\epsilon$ should be chosen, we first compare the power of the Oracle decision function to the usual simple decision function.  The uniformly most powerful unbiased simple decision function, which does not split the data set but makes use of $W_m =
Y_m+Z_m$ as test data, is defined via
$$\delta_m^{(s)}(W_m;\eta_m) = I\left(W_m\leq l_m^{(s)}(\eta_m)\right) + I\left(W_m\geq u_m^{(s)}(\eta_m)\right),$$
where $l_m^{(s)}(\eta_m) = \Phi^{-1}(\eta_m/2)$ and $u_m^{(s)}(\eta_m) = \Phi^{-1}(1-\eta_m/2)$.  The power of this simple decision function is
\begin{eqnarray*}\beta_m^{(s)}(\mu_m, \eta_m) = \Phi\left(l_m^{(s)}(\eta_m) - \mu_m\right) + 1 - \Phi\left(u_m^{(s)}(\eta_m) - \mu_m\right).
 \end{eqnarray*}
From expression (\ref{orpow}) and the definition of $\delta_m^{(or)}(\bm{\mu},Z_m;\eta_m)$, the power of the Oracle decision function is
\begin{eqnarray*}
\lefteqn{\beta_m^{(or)}(\mu_m,\lambda,\eta_m)}&&\\
&&= \Phi\left(l_m^{(or)}(\mu_m,\eta_m) - \sqrt{1-\lambda^2}\mu_m\right)
 + 1-\Phi\left(u_m^{(or)}(\mu_m,\eta_m) - \sqrt{1-\lambda^2}\mu_m\right)
\end{eqnarray*}
The potential gain in power of the Oracle decision function over the simple decision function comes from the refinement of the upper-tail and lower-tail cutoffs. For example, suppose $\mu_m = -1$, $\eta_m = .05$, and $\lambda^2 = 0$.  Then, $l_m^{(or)}(-1,.05) = -1.645$ and $u_m^{(or)}(-1,.05) = \infty$, while $l_m^{(s)}(.05) = -1.96$ and $u_m^{(s)}(.05) = 1.96$.  Hence,
$\beta_m^{(or)}(-1, 0, .05) = \Phi(-1.645 + 1)$, while $\beta_m^{(s)}(-1,.05) = \Phi(-1.96 + 1) + [1-\Phi(1.96+1)]\approx \Phi(-1.96+1)$.  The Oracle decision function power is then larger than the simple decision function power since its lower-tail cutoff is -1.645 rather than -1.96.

\begin{figure}[!h]
\centering
\epsfig{file=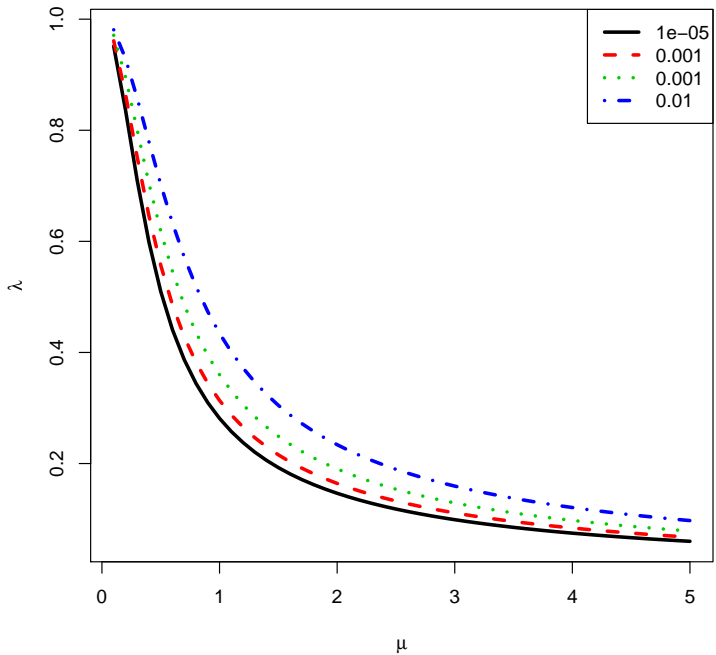, width=4in}
\caption{The region $\left\{(\mu_m,\lambda^2): \beta_m^{(or)}(\mu_m, \lambda, \eta_m)>\beta_m^{(s)}(\mu_m,\eta_m)\right\}$ for
$\eta_m = .01, .001, .0001, .00001$ is the area to the left of each curve.\label{region}}
\end{figure}

However, to implement the Oracle decision function, we must take $\lambda^2>0$ so that some data can be used to estimate the Oracle cutoffs.  The potential loss in power as a result of only using $(1-\lambda^2)100$\% of the data as test data is manifested in the decreased Oracle effect size $|\sqrt{1-\lambda^2}\mu_m|$.  For example, when $\mu_m = -1$ and $\lambda^2 = .4$, then the effect sizes of the Oracle and simple decision functions are .6 and 1, respectively, and the resulting powers are approximately $\Phi(-1.96 + 1) = \Phi(-.96)$ and $\Phi(-1.645 +.6) = \Phi(-1.045)$, respectively.  Hence, the refined cutoffs of the Oracle decision function could not compensate for the decreased effect size, and as a consequence the compound decision function will be less powerful than the simple decision function. We more thoroughly examine this notion using Figure \ref{region}, which depicts the regions of $\left\{(\mu_m,\lambda^2)\right\}$ where $\beta_m^{(or)}(\mu_m,\lambda^2,\eta_m)>\beta_m^{(s)}(\mu_m,\eta_m)$ for several different values of $\eta_m$.  We see that the Oracle decision function power is greater than the simple decision function power for larger values of $\lambda$ when $\mu_m$ is near 0. Hence, the potential gain in power of the compound decision function is more pronounced in the frequently encountered low-power setting.
It is important to emphasize that even if $\lambda^2$ is chosen so that some Oracle decision functions are less powerful than the simple decision function, it may still be the case that the \textit{average} power (computed via expression (\ref{oravepow})) of the Oracle decision functions is larger than the \textit{average} power of the simple decision functions.

We now examine the properties of $h_m(Y)$ and the power of the compound decision function.  The ideal setting is that for small $\lambda^2$, $h_m(Y)=I(\mu_m\leq 0)$ with probability 1.  Then, it would follow from the definitions of $\delta_m^{(or)}$ and $\delta_m^{(c)}$ that
\begin{eqnarray*}
\beta_m^{(c)}(\bm{\mu},\lambda^2, \eta_m) &=& E_F\left[E_F\left\{\delta_m^{(c)}(Y,Z_m;\eta_m)|Y\right\}\right] \\
&=&E_F[\delta_m^{(or)}(\bm{\mu},Z_m;\eta_m)] = \beta_m^{(or)}(\bm{\mu},\lambda^2,\eta_m)
\end{eqnarray*}
In Theorem \ref{asymptotic}, we see that this ideal scenario is achieved asymptotically (in the number of tests $M$) under the two-group model for any arbitrary choice of $\lambda^2$ and $\epsilon$.  See \cite{Efr08} for a discussion regarding this type of model, and \cite{GenWas02}, \cite{Sto03}, \cite{JinCai07}, \cite{RomWol07}, \cite{SunCai07}, among others, for other interesting asymptotic results in this two-group setting.  Below, since we will let the number of tests $M$ tend to $\infty$, we write $\bm{Y}_M\equiv Y$ and $\bm{J}_M \equiv J$ to indicate that the vectors have length $M$, and the notation ``$\stackrel{d}\rightarrow$'' and ``$\stackrel{p}\rightarrow$'' means ``converges in distribution'' and ``converges in probability'', respectively.
\begin{theorem} \label{asymptotic}
Suppose that $E[\bm{Y}_M|\bm{J}_M] = \lambda^2\bm{\mu}_M$ with $\bm{\mu}_M = \theta \bm{J}_M$ for some nonzero scalar $\theta$ and $\bm{J}_M$ a vector of independent and identically distributed Bernoulli random variables with success probability $p\in(0,1]$, and that $Cov(\bm{Y}_M|\bm{J}_M) = \lambda^2\bm{I}_M$.  Suppose further that estimators of $\theta$ and $\tau$ in expression (\ref{hhat}) are defined as in expression (\ref{estimators}) and that
$$S^2(\bm{Y}_M) \stackrel{p}\rightarrow E[S^2(\bm{Y}_M)] = \lambda^2 + \lambda^4\theta^2p(1-p)$$ as $M\rightarrow \infty$, where $S^2(\bm{Y}_M)$ is the sample variance of $\bm{Y}_M$.  Then for any $\epsilon > 0$ and $\lambda^2\in(0,1]$,
$$h_m(\bm{Y}_M)\stackrel{p}\rightarrow I(\theta\leq 0)$$
and
$P_{\Delta_m^{(c)}}(\bm{Y}_M,Z_m)\stackrel{d}\rightarrow P_{\Delta_m^{(or)}}(\mu_m,Z_m)$
as $M \rightarrow \infty$.
\end{theorem}
Several important points should be made.   First, Theorem \ref{asymptotic} holds for any fixed $\epsilon>0$, and hence, at least for large $M$ and under the two-group model, the choice of $\epsilon$ becomes less of an issue.  It should also be noted that $\bm{Y}_M$ need not have a multivariate Normal distribution.  It is only necessary that $S^2(\bm{Y}_M)$ consistently estimate the marginal variance of $Y_m$.  
Finally, the compound $p$-value is $\mathcal{F}_{M_0}^N$-uniform and independent regardless of $M$.
In the next subsection, we study the performance of the compound $p$-value when the two-group model is not satisfied, and $h_m(\bm{Y}_M)$ need not estimate $I(\mu_m\leq 0)$ well.



\subsection{Simulation Study}
In this section, we compare the performance of the compound, Oracle, and simple $p$-values in terms of their ability to allow for multiple testing procedures to be more powerful.  In particular, we consider the BH procedure in \cite{BenHoc95} and the $Q$-value procedure in \cite{Sto02} and \cite{Sto03}.  The procedures are defined as follows.  Let $\bm{p} = (p_m, m\in\mathcal{M})$ be a collection of $p$-values for testing $H_{m0}$ vs. $H_{m1}$ for
$m\in\mathcal{M}$, and denote the ordered $p$-values by $p_{(1)}\leq p_{(2)}\leq . . . \leq p_{(M)}$.  For each pair of
hypotheses, the BH decision function is $\delta_{m,BH}(\bm{p} ;\alpha) = I(p_m\leq \alpha
J_{BH}(\bm{p})/M)$ where
$$
J_{BH}(\bm{p}) = \max\left\{m\in\mathcal{M}:p_{(m)}\leq \alpha \frac{m}{M}\right\}.
$$
The $Q$-value decision function is defined via $\delta_{m,Q}(\bm{p};\alpha) = I(\hat q_m(\bm{p})\leq \alpha)$, where $\hat q_m(\bm{p})$ is the estimated $q$-value for the $m$th pair of hypotheses, defined via
$$\hat q_m(\bm{p}) = \inf_{\gamma\geq p_m} \widehat{pFDR}(\gamma).$$
Here, $\widehat{pFDR(\gamma)}$ is the estimated positive False Discovery Rate ($pFDR = E[V/R|R>0]$) incurred by rejecting all null hypotheses with a $p$-value less than or equal to $\gamma$.  Hence, the $q$-value can be thought of as the smallest possible $pFDR$ allowing for the rejection of $H_{m0}$.  Estimates of the $pFDR$ proposed in \cite{Sto02}, which were shown to be conservative in certain settings, are obtained using the R package {\tt q-value}.  See \cite{Sto02} for more details.

The important point is that the $Q$-value procedure is designed to control the $pFDR$ at level $\alpha$ assuming that $p$-values are independent and uniformly distributed under the null hypotheses.  Likewise, \cite{BenHoc95} show that the BH procedure controls the $FDR = E[V/R|R>0]\Pr(R>0)$ at level $\alpha\frac{M_0}{M}\leq \alpha$ under the independence and uniformity assumptions.  Since the simple, Oracle, and compound $p$-values developed in this paper are all $\mathcal{F}_{\mathcal{M}_0}^N$-uniform and -independent, both procedures are valid when applied to any of these $p$-values.

In our simulation, we considered the same model and hypotheses as in the last section with $\mathcal{M} = \{1,2, ..., 5000\}$, $\mathcal{M}_1 =\{1, 2, ..., 1000\}$, and $\mathcal{M}_0 = \{1001, 1002, ..., 5000\}$.  For $\mathcal{M}_0 = \{1001, 1002, ..., 5000\}$, $\mu_m = 0$.  Hence, 20\% of null
hypotheses are false.  For $m\in\mathcal{M}_1$, we take $\mu_m =
\Phi^{-1}(m/1001;\theta,\tau)$, where $\Phi^{-1}(\cdot;\theta,\tau)$ is the quantile function for a normal distribution with mean $\theta$ and variance $\tau^2$.  Hence, the $\mu_m$s are the expected values of the order statistics from a normal distribution with mean $\theta$ and variance $\tau^2$, thereby allowing the location and spread of the signal, under the alternative hypotheses, to be governed by $\theta$ and $\tau$.  Here, we will consider all combinations of $\theta\in \{0, 2, 4\}$ and $\tau = \{0, 2\}$.  Notice that when $\theta = 0$, the $\mu_m$s from false null hypotheses are symmetric about 0.  \cite{SunCai07} showed that in this setting, and under a two-group model, simple $p$-values tend to yield efficient multiple testing procedures.  When $\theta$ is not 0, however, the signals are not symmetric about 0.  Also, when $\tau = 0$, the two-group model is satisfied and Theorem 4 is applicable.  When $\tau = 2$, the two-group model is not satisfied, and it need not be the case that $I(\mu_m\leq 0)$ is ``well-estimated'' by $h_m(y)$.  For the $k$th replicated data set, vectors of training data and test data are generated according to $Y_k\sim
MVN(\lambda^2\bm{\mu},\lambda^2\bm{I})$ and $Z_k\sim MVN((1-\lambda^2)\bm{\mu},(1-\lambda^2)\bm{I})$, respectively.  For $k=1, 2, ..., K=1000$, both procedures are applied to the collection of Oracle $p$-values computed as in (\ref{OracleP-val}), and three different collections of compound $p$-values in (\ref{CompP-val}) computed by taking $\hat p(y;\lambda)$, $\hat p(y;2\lambda)$, and $\hat p\equiv 1$.  The choice of $\epsilon = \lambda$ and $2\lambda$, which is 1 and 2 standard deviations of $Y_m$ under $H_{m0}$, was recommended in \cite{Efr04} for this type of estimator.  The usual simple $p$-values, which make use of all of the data $W_m = Y_m + Z_m$ as test data rather than just $Z_m$, are computed via $P_{\Delta_m^{(s)}}(W_m) = 2[1-\Phi(|W_m|)]$.

Both procedures were applied to all types of $p$-values for all data sets at $\alpha = .05$.  The average sample $pFDR$ of the $Q$-value procedure was less than .05 for all configurations and $p$-value types.  Similarly, the average sample $FDR$ of the BH procedure was less than .05 for all configurations and $p$-value types.   The average power of the BH procedure for a particular set of $p$-values and $(\theta,\tau)$-combination is estimated via
$$\hat \beta = \frac{1}{K}\sum_{k=1}^{K}\left[\frac{1}{M_1} \sum_{m\in\mathcal{M}_1}\delta_{m,BH}(\bm{p}_k;\alpha)\right].$$
The average power of the $Q$-value procedure is computed analogously.
Results are presented in Table \ref{tablesimul}.

\begin{table}
\caption{\label{tablesimul}The average power of the BH and $Q$-value procedures when making use of simple $p$-values ($\lambda^2 = 0$), Oracle $p$-values, and compound $p$-values where $p$ is estimated with $\hat p(y;\lambda)$, $\hat p(y;2\lambda)$, and
1.}
\centering
\fbox{
\begin{tabular}{ccc||cccccccccc}  &&&\multicolumn{10}{c}{BH Procedure}\\ \hline
 & & && \multicolumn{3}{c}{$\tau=0$}&&\multicolumn{5}{c}{$\tau=2$} \\   \cline{5-7} \cline{9-13}
$\lambda^2$&&  && $\theta=2$&&$\theta=4$&&$\theta=0$&&$\theta=2$&&$\theta=4$\\  \hline \hline
\bm{0}&& Simple && 0.10 && 0.92 && 0.16 && 0.36 && 0.72 \\ \hline
.01&&Oracle && 0.18 && 0.95 && 0.20 && 0.40 && 0.76 \\
.01&& p=1 && 0.15 && 0.94 && 0.13 && 0.37 && 0.74 \\
.01&& $\hat p(y;.01)$ && 0.18 && 0.95 && 0.10 && 0.38 && 0.76 \\
.01&& $\hat p(y;.02)$&& 0.18 && 0.95 && 0.09 && 0.38 && 0.76 \\
 \hline

.05&& Oracle && 0.16 && 0.94 && 0.19 && 0.39&& 0.75 \\
.05&&  p=1 && 0.12 && 0.93 && 0.15 && 0.36 && 0.73 \\
.05&&  $\hat p(y;.05)$ && 0.16 && 0.94 && 0.13 && 0.37 && 0.75 \\
.05&&  $\hat p(y;.1)$ && 0.16 && 0.94 && 0.12 &&0.37 && 0.75 \\
\hline

.10 &&Oracle && 0.14 && 0.93 && 0.17 && 0.38 && 0.74 \\
 .10 && p=1 && 0.10 && 0.92 && 0.14 && 0.35 && 0.72 \\
.10 &&  $\hat p(y;.1)$ && 0.14 && 0.93 && 0.15 && 0.36 && 0.74 \\
.10 && $\hat p(y;.2)$ && 0.14 && 0.93 && 0.14 && 0.36 && 0.74 \\
\hline
.20&& Oracle && 0.10 && 0.89 && 0.15 && 0.34 && 0.71 \\
.20&& p=1 && 0.07 && 0.88 && 0.12 && 0.32 && 0.70 \\
.20&& $\hat p(y;.2)$&& 0.10 && 0.89 && 0.13 && 0.33 && 0.71 \\
.20&&  $\hat p(y;.4)$ && 0.10 && 0.89 && 0.13 && 0.33 && 0.71 \\   \hline \hline

&&&\multicolumn{10}{c}{Q-value Procedure}\\ \hline

\bm{0}&& Simple && 0.12 && 0.93 && 0.16 && 0.37 && 0.74 \\ \hline
.01&& Oracle && 0.22 && 0.96 && 0.21 && 0.42 && 0.77 \\
.01&& p=1 && 0.18 && 0.95 && 0.13 && 0.38 && 0.75 \\
.01&& $\hat p(y;.01)$ && 0.22 && 0.96 && 0.10 && 0.39 && 0.77 \\
.01&&  $\hat p(y;.02)$ && 0.22 && 0.96 && 0.10 && 0.39 && 0.77 \\
 \hline

.05&& Oracle && 0.20 && 0.95 && 0.20 && 0.41 && 0.76 \\
.05&&  p=1 && 0.15 && 0.94 && 0.15 && 0.37 && 0.74 \\
.05&&  $\hat p(y;.05)$ && 0.20 && 0.95 && 0.14 && 0.38 && 0.76 \\
.05&&  $\hat p(y;.1)$ && 0.20 && 0.95 && 0.12 && 0.38 && 0.76 \\
\hline

.10&& Oracle && 0.17 && 0.94 && 0.19 && 0.39 && 0.75 \\
.10&&  p=1 && 0.13 && 0.93 && 0.15 && 0.36 && 0.74 \\
.10&& $\hat p(y;.1)$ && 0.18 && 0.94 && 0.15 && 0.36 && 0.75 \\
.10&& $\hat p(y;.2)$ && 0.18 && 0.94 && 0.15 && 0.36 && 0.75 \\
\hline

.20&& o && 0.13 && 0.91 && 0.16 && 0.36 && 0.73 \\
.20&&  p=1 && 0.09 && 0.90 && 0.13 && 0.34 && 0.71 \\
.20&&  $\hat p(y;.2)$ && 0.13 && 0.91 && 0.14 && 0.34 && 0.72 \\
.20&&  $\hat p(y;.4)$ && 0.13 && 0.91 && 0.14 && 0.33 && 0.72 \\
\end{tabular}
}
\end{table}

First, notice that when $\tau = 0$ and the two-group model is satisfied, the power of a multiple testing procedure which makes use of the Oracle $p$-values is equivalent to the power of the procedure when using compound $p$-values for any choice of $\epsilon$ or $\lambda^2$, just as Theorem 4 predicted. Further, this power can be substantially larger than the power of the same multiple testing procedure that makes use of the simple $p$-values, especially in the low-power setting.  For example, for $\lambda^2 = .01$, $\theta = 2$, and $\tau^2 = 0$, the power of the $Q$-value procedure is increased by 83\% when using the compound $p$-values (when using $\hat p(Y;\epsilon)$) over the simple $p$-values (.22/.12 = 1.83).  The power of the Q-value procedure is increased by 80\% (.18/.1 = 1.8). This supports findings in the previous subsection (see Figure 1), where it was argued that the greatest potential for gain in power occurs when $\mu_m$ is near 0.

Likewise, as discussed in the previous subsection, when too much data is used as training data, Oracle $p$-values, and hence compound $p$-values, need not yield more powerful multiple testing procedures.  For example, when $\lambda^2 = .2$, the average power of the simple decision functions is greater than the average power of the Oracle decision functions in most settings (the exception being in the frequently encountered low power setting when $\theta = 2$ and $\tau = 0$).  This scenario can and should be avoided in practice by choosing $\lambda^2<.2$.

When $\tau^2 = 2$ and $\lambda^2\leq .1$ (note that the two-group model is not satisfied so that $h_m(y)$ need not estimate $I(\mu_m\leq 0)$ well), we see that the compound $p$-values still result in more power than the usual simple $p$-values.  The only exception is the setting when $\theta = 0$.  However, the loss in power in this setting is small relative to the gain in power in the non-symmetric settings, especially when a small portion of data are used as training data and the data from false null hypotheses are highly concentrated.

In general, if less than 10\% of the data is being used as training data, compound $p$-values will tend to lead to more powerful multiple testing procedures.  The biggest gain in power occurs in the low-power setting when the signals (the $\mu_m$s) are identical.  As the signals become more dispersed, less power is gained.

\subsection{Comparison to Other Compound Methods}
The sample splitting approach allows for more modeling assumptions regarding the joint behavior of the data, and at the same time enjoys a certain robustness property.  To see why, first a discussion regarding relaxing assumptions from the previous sections is provided.  Then, the methodology is compared to competing strategies.

In general, one may compute a test statistic for test data via $T_m= \mathcal{T}(X[m,\bar{T}])$, where $\mathcal{T}$ is some test statistic so that under $H_{m0}$, $T_m \sim F$.  Then, $Z_m = \Phi^{-1}(F(T_m))$ has standard normal distribution (so long as $F$ is continuous) under the null hypothesis by the probability integral transformation.  Compound $p$-values can then be computed as in the previous section (with $\lambda^2 = 1$).  This is demonstrated in detail in the following section.  Then, from Theorem \ref{mainthm}, the resulting compound $p$-values will be uniformly distributed under $H_{m0}: T_m\sim F$.  If test data are independent under the null hypotheses, $p$-values will remain independent under the null hypotheses as well.  Hence, \textit{regardless of the distribution of the test statistics under the alternative hypothesis}, the applied multiple testing procedure, whichever is chosen, will be valid.  It is only necessary that the appropriate test statistic be chosen so that $T_m$ does indeed have distribution function $F$ under $H_{m0}$.  For robust test statistics for multiple testing procedures see \cite{HabPen11a}.

To better understand the sample splitting approach, it is useful to first discuss procedures based on the two-group model.  \cite{EfrTib01}, \cite{SunCai07}, among others, assume that $Z_m\sim f = p f_0 + (1-p) f_1$ where $f_0$ is the density of $Z_m$ under $H_{m0}$, $f_1$ the density of $Z_m$ under $H_{m1}$, and $p$ is a mixing proportion.  \cite{SunCai07} show that the Lfdr statistic, defined $$\widehat{Lfdr}(z_m) = \frac{\hat{p}f_0(z_m)}{\hat p f_0(z_m) + (1-\hat{p})\hat f_1(z)_m}$$
can be used to control the FDR (asymptotically in $M$) so long as $p\in(0,1)$ and $\hat p $ and $\hat f_1$ are consistent estimators.  Since the validity of the procedure requires consistent estimation of $f_1$, it is vital that a flexible model for $f_1$ be utilized, as is done in the above references.  Added efficiency stems from the fact that the Lfdr statistic is proportional to the estimated likelihood ratio statistic $\hat \Lambda(z_m) = \frac{\hat f_1}{f_0}(z_m)$.  See \cite{Hab11} for details.  The procedure is \textit{compound} because joint behavior of the data is utilized, i.e. information is pooled, through the estimation of $f_1$ with $z_1, z_2, ..., z_M$.  The resulting decision rule, which can be written $\bm{\delta}(\bm{z}) = [I(\hat\Lambda(z_1) > c),..., I(\hat\Lambda(z_M)>c)]$ for some cutoff $c$, is referred to as \textit{symmetric} since for all permutation operators $\bm{\tau}$, $\bm{\tau}(\bm{\delta}(\bm{z})) = \bm{\delta}(\bm{\tau}(\bm{z}))$.

In our example in Section 4, we allowed for data to vary according to a different distribution under each alternative hypothesis.  Specifically it was assumed that $Z_m \sim f = p f_0 + (1-p)f_m$, where $f_m$ is an unknown normal density with mean $\mu_m$.  The result was a compound decision rule that depended upon $M$ different likelihood ratio statistics $\hat \Lambda_m(z_m) = \frac{\hat f_m}{f_0}(z_m),m\in\mathcal{M}$, and hence need not be \textit{symmetric}.  We focused on the estimation of $I(\mu_m<0)$ since the form of the likelihood ratio statistic only depends upon this quantity in the normal setting.  The joint behavior of the data was modeled by assuming that $\mu_m\sim N(\theta,\tau)$, and information is pooled by then allowing $\hat{f}_m$ to depend upon all the training data via $\hat{\theta}(y)$ and $\hat{\tau}(y)$.  \cite{Sto07} also considered basing decision rules on $M$ different normal models.

The main difference between our approach and the aforementioned is that the information pooling is done using only training data, rather than all of the data, and that $p$-values for each decision function are provided.  This sample spitting approach allows for \textit{valid} $p$-values, even if the data are incorrectly modeled under the alternative hypothesis, and even if the number of tests $M$ is small.  For this reason, it is reasonable to base each Oracle decision rule on stronger modeling assumptions, as was done here.  Further, by computing $p$-values for each test, any number of multiple testing procedures could be employed to control the error rate of interest, including but not limited to the FDR, pFDR, or FWER.

\section{Application to a Real Data Set}\label{sec5}
In this section, we analyze the microarray data in \cite{Sin02} using methods from the previous two sections. This data was
also analyzed in \cite{Efr09}.  Here, $X[m,n]$ is the $m$th gene expression measurement from the $n$th microarray, where for
$n\in\mathcal{N}_1 = \{1, 2, ..., 50\}$, microarray $n$ is from an individual without prostate cancer and for $n\in
\mathcal{N}_2 = \{51, 52, ..., 102\}$, microarray $n$ is from an individual with prostate cancer.  The goal is to determine
which genes, if any, are differentially expressed across treatment groups.

We assume that $X[m,n]\stackrel {i.i.d.}\sim N(\gamma_m,\sigma_m^2)$ for $n\in \mathcal{N}_1$ and
$X[m,n]\stackrel{i.i.d}\sim N(\gamma_m+\mu_m,\sigma_m^2)$ for $n\in\mathcal{N}_2$.  The $m$th null and alternative
hypotheses are $H_{m0}:\mu_m = 0, F_m\in \mathcal{F}^{Norm}$ and $H_{m1}:\mu_m \neq 0, F_m\in \mathcal{F}^{Norm}$, where
$\mathcal{F}^{Norm}$ is the collection of all normal distribution functions.

\begin{table}
\caption{Depiction of a portion of the microarray data in \cite{Sin02}, where $x[m,n]$ is the $m$th gene expression level from the $n$th individual. Data for the $n$th microarray is $x[,n]$ and data for the $m$th gene is $x[m,]$ \label{Sindata} }
\centering
\fbox{
\begin{tabular} {c|cccc|cccc}
&\multicolumn{4}{|c|}{control group} & \multicolumn{4}{|c}{cancer group}\\
&$x[,1]$&$x[,2]$&...&$x[,50]$&$x[,51]$&$x[,52]$&...&$x[,102]$\\ \hline
$x[1,]$&-.931&-.840&...&3.81&-1.12&1.01&...&-.001 \\
$x[2,]$&-1.07&-.880&...&-.477&-.571&-.811&...&-.836  \\
$\vdots$ & $\vdots$&$\vdots$ &$\ddots$&$\vdots$&$\vdots$&$\vdots$&$\ddots$&$\vdots$\\
$x[6033,]$& -.754 & -.708&... & -.011 & .457 & .578 & ... & -.162  \\ \hline
\end{tabular}
}
\end{table}

We present the form of the compound and simple $p$-value statistics.  Here, $T = T_1\cup T_2$, where $T_1$ and $T_2$ index training data from control and treatment groups, respectively, and $\bar T = \bar T_1\cup \bar T_2$, where $\bar{T}_1 = \mathcal{N}_1\setminus T_1$ and $\bar{T}_2 = \mathcal{N}_2\setminus T_2$ index test data from control and treatment groups, respectively.  For this data, since the simulation studies from the previous section suggest that between 1 and 10 percent of data should be used as training data, we (randomly) select 4 of our 102 microarrays as training data ($T_1 = \{10, 22\}$ and $T_2 = \{60, 88\}$). The two sample $T$-test statistic for $H_{m0}$ based on test data $X[m,\bar T]$ is
\begin{equation}\nonumber \label{T-test}\mbox{T}_m(X[m,\bar T]) =
\frac{\sum_{n\in \bar T_2}X[m,n]/n_{\bar T_2} - \sum_{n\in \bar T_1}X[m,n]/n_{\bar T_1}}{s_{pm}\sqrt{\frac{1}{n_{\bar T_1}} + \frac{1}{n_{\bar T_2}}}}
\end{equation}
where $n_{A} = |A|$ and $s_{pm}$ is the pooled sample standard deviation of $X[m,\bar T_1]$ and $X[m,\bar T_2]$.  To remain consistent with notation in the previous sections, we transform $\mbox{T}_m$ via $Z_m = \Phi^{-1}(\mathcal{T}_{n_{\bar T} - 2}(\mbox{T}_m(X[m,\bar T]))$ so that $Z_m\sim N(0,1)$ under $H_{m0}$ by the probability integral transformation.  In a similar fashion, we transform the training data via $Y_m = $ \\$\Phi^{-1}(\mathcal{T}_{n_{T}-2}(\mbox{T}_m(X[m,T])))$, where $\mbox{T}_m(X[m,T])$ is Student's two-sample $T$-test as above but computed on $X[m,T_1]$ and $X[m,T_2]$.  It is important to note that since $\lambda^2$ is now fixed, we do not parameterize our test data and training data to have mean and variance that depends on $\lambda^2$. The compound decision function can then be defined via
\begin{eqnarray*}\nonumber \delta_m^{(c)}(Y,Z_m;\eta_m) = \left\{
\begin{array}{l l} \nonumber
1 & \mbox{ if } Z_m \leq  \Phi^{-1}(\eta_m h_m(Y))\nonumber \\
1& \mbox{ if } Z_m \geq \Phi^{-1}(1-\eta_m[1- h_m(Y)]) \nonumber\\
0 & \mbox{otherwise},
\end{array} \right.
\end{eqnarray*}
where $Y = (Y_1, Y_2, ..., Y_M)$.  It can be verified using arguments from Section 4 that the compound $p$-value can be written as in expression (\ref{CompP-val}), and that $h_m(Y)$ should estimate $I(\mu_m\leq 0)$.  Hence, we define $h_m(Y)$ as in (\ref{hhat}) with $\lambda^2 = 1$ since $Var(Y_m) = 1$.  For the compound $p$-values, we consider taking $\epsilon = 1$ and $2$ in $\hat{p}(y;\epsilon)$ since this corresponds to 1 and 2 standard deviations of $Y_m$ under $H_{m0}$.  We also take $\hat p = .1$ as in \cite{Efr09} and $\hat p = 1$ as in the previous section.  The usual two sample $T$-test $p$-values were computed via $P_{\Delta_m^{(s)}}(X[m,]) = 2[1-\mathcal{T}_{100}(|T(X[m,])|)$, where $\mbox{T}(X[m,])$ is the two sample $T$ test statistic as above but with $\bar T_1 = \mathcal{N}_1$ and $\bar{T}_2 = \mathcal{N}_2$ indexing all of the data from control and treatment groups.

\begin{figure}[h!]\centering
\epsfig{file=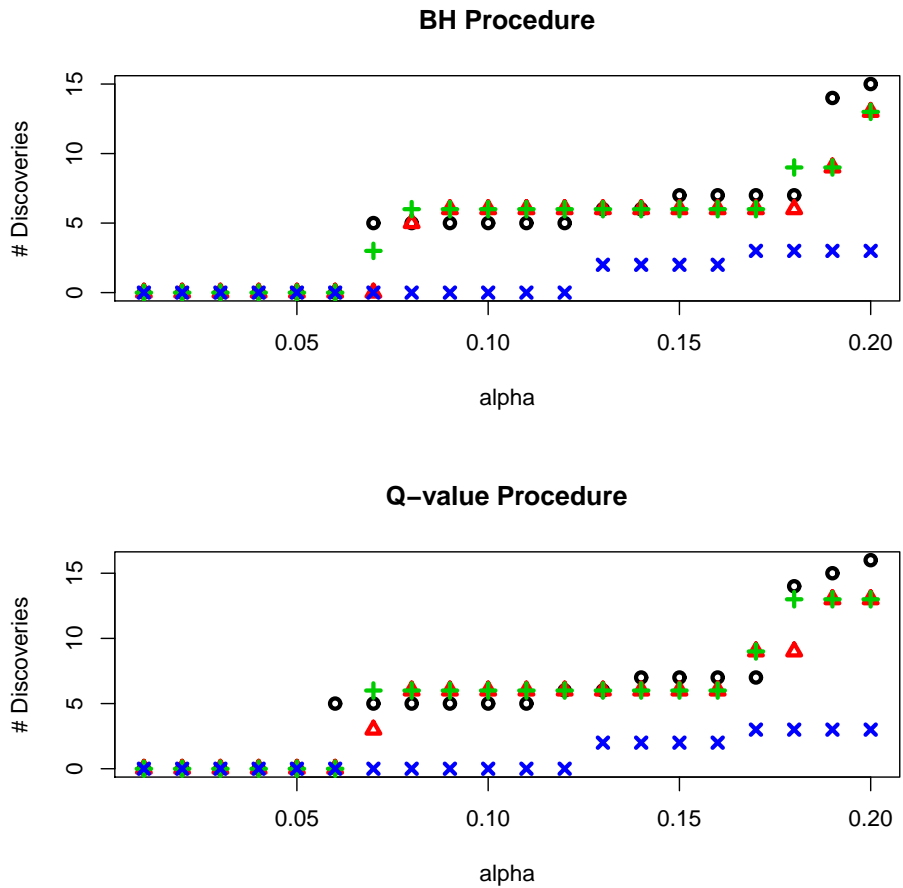,width=5in}
\caption{\label{app} The number of discoveries when applying the BH (top) and Q-value (bottom) procedures to simple $p$-values ({\color{blue}{x}}) and compound $p$-values when $p$ is estimated with $\hat p(y;2)$ (o), assumed to be $.1$ ({\color{red}$\triangle$}), and assumed to be 1 ({\color{green}+}). }
\end{figure}
The number of discoveries made by the BH and $Q$-value procedures when applied to each of the different collections
of $p$-values at levels $\alpha = .01, .02, ..., .2$ are presented in Figure \ref{app}. Results when compound $p$-values
made use of $\hat p(y;1)$ are not presented because we get a negative estimate of $p$.  Such estimates are not
uncommon when $p$ and $\epsilon$ are near 0 due to the fact that the bias of $\hat p(Y;\epsilon)$ is negligible in this setting.  See \cite{Efr04} for a discussion regarding this issue.
We see that when making use of any of the compound $p$-values, rather than the simple $p$-values, both procedures always make at least as many or more (sometimes substantially more) discoveries.  For example, when the BH procedure is applied at $\alpha = .2$ to compound $p$-values with $\hat p(y;2) = .017$, 15 discoveries, rather than 3, are made.  For $\alpha = .1$, the compound $p$-values which assume $p=.1$ and $p=1$ allow for the BH procedure to make 5 and 6 discoveries, respectively, while the use of the simple $p$-values leads to 0 discoveries.  Results are similar for the Q-value procedure in that compound $p$-values always allow for at least as many discoveries, and sometimes allow for substantially more discoveries.

\section{Concluding Remarks}
Recent multiple testing research has established that compound multiple testing procedures are typically more efficient than simple multiple testing procedures.  In this paper, we have shown that these multiple testing procedures can be made even
more efficient by making use of compound test statistics.  We have limited our study to compound $p$-value statistics,
largely due to the fact that a substantial number of multiple testing procedures make use of $p$-value statistics, thus
making results in this paper widely applicable.

Here, the data were split into training and test data, and only training data (as opposed to all the data), were utilized to borrow information across tests.  The main advantage of this data-splitting approach over the usual double dipping approach is that validity of the resulting $p$-values and multiple testing procedure is guaranteed, even if data are poorly modeled under the alternative hypotheses, and even for a small number of tests $M$. Intuition suggests that the disadvantage of this approach is that in some settings efficiency will be sacrificed since less data is utilized to estimate parameters governing the form of the Oracle decision rule.  A more thorough comparison of this approach and the double dipping approach is warranted, but is beyond the scope of this paper.  See also \cite{Pen11} for a discussion on this issue.

The examples in this paper could likely be improved upon by considering other types of models for the joint behavior of the data, as well as other type of estimators.  Method of moment estimators were utilized to allow for easy-to-compute $p$-values.

The assumption that test statistics are independent under the null hypotheses may not be satisfied in practice.  In this setting, we cannot expect compound or simple $p$-value statistics to be independent under the null hypotheses.  However, many $p$-value based multiple testing procedures, including some of those mentioned in the Introduction, do not require the independence condition to be satisfied. Results in Sections 2 and 3 can still be used to develop compound $p$-value statistics satisfying the uniformity condition, which can then be used in these multiple testing procedures.  See \citet{BenYek01};\cite{Sar02,Sar07,SunCai09} for more on relaxing the independence condition.

In closing, we reiterate the intent in this paper is not to develop a new compound multiple testing procedure, but rather
to develop compound $p$-value statistics for use in existing multiple testing procedures.  We have only studied the effects
of compound $p$-value statistics on two compound multiple testing procedures, but we suspect that most multiple testing
procedures will behave in a more efficient manner if they are used in conjunction with compound, rather than simple, $p$-value statistics.

\section{Appendix: Proofs}
\indent \hspace{.15in} \textbf{Proof of Theorem \ref{thmPvsD}:} It suffices to show that $P_{\Delta_m}(X)$ is $\mathcal{F}_{m0}$-uniform for every $m\in\mathcal{M}_0$.
But since $\sup_{F\in\mathcal{F}_{m0}} E_F(\delta_m(X;\eta_m)) = \eta_m$ for every $\eta_m\in [0, 1]$, the result follows
from Theorem 2.3 in \cite{HabPen11a} by taking $X_m=X$.

\textbf{Proof of Theorem \ref{thmindependent}:}
Suppose we could show that $\Pr_F(\delta_m(X;t_m) = I(P_{\Delta_m}(X)\leq t_m)) = 1$ for every $F\in\mathcal{F}$,
$t_m\in[0,1]$, and $m\in\mathcal{M}$.  Then it will follow that
\begin{eqnarray*}
\lefteqn{\Pr_F\left(\bigcap_{m\in \mathcal{M}}[\delta_m(X;t_m) = I(P_{\Delta_m}(X)\leq t_m)]\right)}\\
 &&= 1 - \Pr_F\left(\bigcup_{m\in \mathcal{M}}[\delta_m(X;t_m) \neq I(P_{\Delta_m}(X)\leq t_m)]\right)\\
&&\geq 1 - \sum_{m\in \mathcal{M}}\Pr_F\left(\delta_m(X;t_m) \neq I(P_{\Delta_m}(X)\leq t_m)\right)\\
&&= 1 - 0 = 1,
\end{eqnarray*}
which will imply that $\Pr_F\left(\bigcap_{m\in\mathcal{M}} [P_{\Delta_m}(X)\leq t_m]\right) =
\Pr_F\left(\bigcap_{m\in\mathcal{M}}[\delta_m(X;t_m)=1]\right).$ The result will then follow from equations
(\ref{independentP}) and (\ref{independentD}).  Therefore, it suffices to show that $\Pr_F(\delta_m(X;t_m) =
I(P_{\Delta_m}(X)\leq t_m)) = 1$.

Fix $F\in\mathcal{F}$.  There exists a null set $N\subset\mathcal{X}$ such that for every $x\in N^c$, $t_m\mapsto \delta_m(x;t_m)$ is right-continuous and nondecreasing with $\Pr_F(X\in N^c) = 1$.  Fix an
$x\in N^c$.  If $a\in \{t_m:\delta_m(x;t_m) = 1\}$, then $\inf\{t_m:\delta_m(x;t_m)=1\}\leq a$ implying that
$P_{\Delta_m}(x)\leq a$.  Hence, $\{t_m:\delta_m(x;t_m) = 1\} \subseteq \{t_m:P_{\Delta_m}(x)\leq t_m\}$ by Definition
\ref{P-value}.  Next, suppose that $a\in\{t_m:P_{\Delta_m}(x)\leq t_m \}$.  Since $\delta_m(x;t_m)$ is right-continuous and
nondecreasing, $\delta_m(x;a) = 1$, so that $a\in \{t_m:\delta_m(x;t_m) = 1\}$ and $\{t_m:\delta_m(x;t_m) = 1\} \supseteq
\{t_m:P_{\Delta_m}(x)\leq t_m\}$. That is, $\delta_m(x;t_m) = I(P_{\Delta_m}(x)\leq t_m)$ for every $x\in N^c$.  Since
$\Pr_F(N^c)=1$, it follows that $P_F(\delta_m(X;t_m)=I(P_{\Delta_m}(X)\leq t_m)) = 1$.

\textbf{Proof of Theorem \ref{mainthm}:}
Theorem \ref{thmPvsD} ensures that $P_\bm{\Delta}(Y,Z)$ is $\mathcal{F}_{\mathcal{M}_0}$-uniform since $\bm{\Delta}$ is a
decision process.  From Theorem \ref{thmindependent}, if $\bm{\Delta}$ is $\mathcal{F}_{\mathcal{M}_0}$-independent, then
$P_\bm{\Delta}(Y,Z)$ is $\mathcal{F}_{\mathcal{M}_0}$-independent.  Hence, it suffices to show that
\begin{eqnarray*}
\lefteqn{\Pr_F(\cap_{m\in\mathcal{M}}[\delta_m(Y,Z_m;\eta_m) = d_m])}\\
&& = \Pr_F(\cap_{m\in\mathcal{M}_1}[\delta_m(Y,Z_m;\eta_m) = d_m])\prod_{m\in\mathcal{M}_0}\Pr_F(\delta_m(Y,Z;\eta_m) = d_m).\end{eqnarray*}
But, since $\Pr_F(\delta_m(Y,Z_m;\eta_m)=d_m|Y) = k_m(\eta_m)$ for  $m\in\mathcal{M}_0$, where $$k_m(\eta_m) = \eta_mI(d_m=1)
+ (1-\eta_m)I(d_m=0),$$ then by the conditions of the theorem and using the laws of iterated expectations, we get
\begin{eqnarray*}
\lefteqn{\Pr_F\left(\bigcap_{m\in\mathcal{M}}[\delta_m(Y,Z_m;\eta_m)=d_m]\right)=  E_F\left\{\Pr_F\left(\bigcap_{m\in\mathcal{M}} [\delta_m(Y,Z_m;\eta_m)=d_m]|Y\right)\right\}} \\
&&=E_F\left(\Pr_F\left(\bigcap_{m\in\mathcal{M}_1}[\delta_m(Y,Z_m;\eta_m)=d_m]|Y\right)\left(\prod_{m\in\mathcal{M}_0}\Pr_F(\delta_m(Y,Z_m;\eta_m)=d_m)|Y\right)\right)\\
&&=  E_F\left(\Pr_F\left(\bigcap_{m\in\mathcal{M}_1}[\delta_m(Y,Z_m)=d_m]|Y\right)\right)\prod_{m\in\mathcal{M}_0}k_m(\eta_m)\\
&&= \Pr_F\left(\bigcap_{m\in\mathcal{M}_1}[\delta_m(Y,Z_m;\eta_m)=d_m]\right)\prod_{m\in\mathcal{M}_0}\Pr_F(\delta_m(Y,Z_m;\eta_m) = d_m).
\end{eqnarray*}

\textbf{Proof of Corollary \ref{cor}:}
Since Condition 1 is satisfied, by Theorem \ref{mainthm} it is sufficient to show that for every $m\in\mathcal{M}_0$ and
$F\in\mathcal{F}_{\mathcal{M}_0}^{N}$, $E_F[\delta_m^{(c)}(Y,Z_m;\eta_m)|Y] = \eta_m$ for any $\eta_m\in[0,1]$.  But if
$m\in\mathcal{M}_0$,
\begin{eqnarray*}
E_F[\delta_m^{(c)}(Y,Z_m;\eta)|Y] &=& E_F\left[\Phi(l_m(Y,\eta_m)) + 1 - \Phi(u_m(Y,\eta_m))\right]\\
&=&  E_F\left[\Phi(\Phi^{-1}(\eta_m h_m(Y)) + 1-\Phi(\Phi^{-1}(1-\eta_m[1-h_m(Y)]))\right] \\
&=& E_F\left[\eta_m h_m(Y) + 1-(1-\eta_m[1-h_m(Y)])\right]\\
&=& \eta_m[h_m(Y)+1-h_m(Y)] = \eta_m
\end{eqnarray*}
for any $\eta_m\in[0,1]$.

\textbf{Proof of Theorem \ref{asymptotic}:}
First, suppose that $m\in\mathcal{M}_0$.  Then it follows from Theorem \ref{thmPvsD} and the fact that  $E_F(\delta_m^{(c)}(\bm{Y}_M,Z_m;\eta_m)) = \eta_m$ and $E_F(\delta_m^{(or)}(\mu_m,Z_m;\eta_m)) = \eta_m$ for every $\eta_m\in[0,1]$, that $P_{\Delta_m^{(or)}}(\mu_m,Z_m) \stackrel{d}=U\stackrel{d}=P_{\Delta_m^{(c)}}(\bm{Y}_M,Z_m)$ where $\stackrel{d}=$ means ``equal in distribution'' and $U$ is a uniform random variate.  Now, for $m\in\mathcal{M}_1 = \mathcal{M}\setminus\mathcal{M}_0 = \{m:\mu_m = \theta\}$, if $h_m(\bm{Y}_M)\stackrel{p}\rightarrow I(\theta\leq 0)$ as $M\rightarrow \infty$, then the Continuous Mapping Theorem (see, for example, page 19 in \cite{Ser80}) and expressions (\ref{OracleP-val}) and (\ref{CompP-val}) imply that $P_{\Delta_m^{(c)}}(\bm{Y}_M,Z_m)\stackrel{d}\rightarrow P_{\Delta_m^{(or)}}(\mu_m,Z_m)$. Hence, it suffices to show that $h_m(\bm{Y}_M)\stackrel{p}\rightarrow I(\theta \leq 0)$.  To do so, we show that
\begin{equation} \label{theta}
\hat{\theta}(\bm{Y}_M) \stackrel{p}\rightarrow k\theta
\end{equation}
for some $k>0$ and
\begin{equation}\label{tau}
\hat\tau^2(\bm{Y}_M)\stackrel{p}\rightarrow 0,
\end{equation}
since these results, together with the Continuous Mapping Theorem, and writing
$$h_m(\bm{Y}_M) = \Phi\left(\frac{-Y_m}{\sqrt{\lambda^2 + 1/\hat{\tau}^2(\bm{Y}_M)}} - \frac{\hat{\theta}(\bm{Y}_M)}{\hat{\tau}^2(\bm{Y}_M)(\lambda^2\hat{\tau}^2(\bm{Y}_M))}\right),$$ imply $h_m(\bm{Y}_M)\stackrel{p}\rightarrow \Phi(-sign(\theta)\infty) = I(\theta\leq 0).$

To show (\ref{theta}), first note that by the inequality in expression (\ref{ineq1}),
\begin{equation}\label{ineq}
0<E\left[1 - \frac{I(-\epsilon \leq Y_m \leq \epsilon)}{\Phi(\epsilon/\lambda) - \Phi(\epsilon/\lambda)}\right]\equiv p^*<p.
\end{equation}
Hence, by the definition of $p(\bm{Y}_M;\epsilon)$ and the weak law of large numbers (WLLN), $\hat{p}(\bm{Y}_M;\epsilon)\stackrel{p}\rightarrow p^*$.  Similarly, since $Var(Y_m)<\infty$, by the WLLN we have $\bar{Y}_M/(\lambda^2 p)\stackrel{p}\rightarrow \theta$.  Hence,
$$\hat{\theta}(\bm{Y}_M) = \frac{\bar{Y}_M}{\lambda^2 \hat{p}(\bm{Y}_M;\epsilon)} = \left(\frac{\bar{Y}_M}{\lambda^2 p}\right)\left( \frac{p}{\hat{p}(\bm{Y}_M;\epsilon)}\right)\stackrel{p}\rightarrow \theta \frac{p}{p^*}.$$

To show (\ref{tau}), first note that $\hat{\theta}(\bm{Y}_M)^2\stackrel{p}\rightarrow \theta^2p^2/(p^*)^2$ since $g(x)=x^2$ is continuous.  From the continuous mapping theorem and since $p/p^*>1$ and $(1-p^*)> (1-p)$ by the inequality in (\ref{ineq}),
\begin{eqnarray*}
\lambda^2 + \lambda^4\hat\theta(\bm{Y}_M)^2 \hat p(\bm{Y}_M;\epsilon)(1-\hat p(\bm{Y}_M;\epsilon)) &\stackrel{p}\rightarrow&
\lambda^2 + \lambda^4\theta^2 \frac{p^2}{(p^*)^2}p^*(1-p^*) \\
&=&\lambda^2 + \lambda^4\theta^2p\left(\frac{p}{p^*}\right)(1-p^*)  \\&>& \lambda^2 + \lambda^4\theta^2p(1-p).
\end{eqnarray*}
Since $S^2(\bm{Y}_M)\stackrel{p}\rightarrow E[S^2(\bm{Y}_M)] = \lambda^2 + \lambda^4\theta^2p(1-p)$, the above result implies
\begin{eqnarray*}
S^2(\bm{Y}_M) - \left[\lambda^2 + \lambda^4\hat\theta(\bm{Y}_M)^2\hat p(\bm{Y}_M;\epsilon)(1-\hat p(\bm{Y}_M;\epsilon))\right] \stackrel{p}\rightarrow c < 0
\end{eqnarray*}
for some $c$.  Hence,
$$\frac{S^2(\bm{Y}_M) - \left[\lambda^2 + \lambda^4 \hat\theta(\bm{Y}_M)^2\hat p(\bm{Y}_M;\epsilon)(1-\hat p(\bm{Y}_M;\epsilon))\right]}{\lambda^4 \hat p(\bm{Y}_M;\epsilon)}\stackrel{p}\rightarrow \frac{c}{\lambda^4 p^*} < 0$$
so that
$$\hat\tau^2(\bm{Y}_M) = \max \left\{\frac{S^2(\bm{Y}_M) -\left[\lambda^2 + \lambda^4\hat p(\bm{Y}_M;\epsilon)(1-\hat p(\bm{Y}_M;\epsilon))\right]}{\lambda^4 \hat p(\bm{Y}_M;\epsilon)}, 0\right\}\stackrel{p}\rightarrow 0. $$

\section*{Acknowledgements}

The authors wish to thank Professors Wensong Wu, Don Edwards, John Grego, Joshua Tebbs, and Hongmei Zhang.  The authors also acknowledge NSF Grant DMS0805809;
National Institutes of Health (NIH) Grant RR17698; and the Environmental Protection Agency (EPA) Grant RD-83241902-0 to the University of Arizona with subaward number Y481344 to the University of South Carolina.  These grants partially supported this work.  This work is based on a portion of the first author's PhD dissertation at the University of South Carolina.


\end{document}